\documentclass[a4paper,twocolumn]{revtex4}
\usepackage{amsmath,amsbsy,amssymb,graphicx,epsf,color,pifont}
\usepackage[mathscr]{euscript}
\usepackage{fouriernc}

\def\babc{\begin{subequations}}
\def\eabc{\end{subequations}}
\def\be{\begin{equation}}
\def\ee{\end{equation}}
\def\ba{\begin{array}}
\def\ea{\end{array}}
\def\nn{\nonumber}
\def\hh{\hspace*{0.25mm}}
\def\h{\hspace*{0.5mm}}
\def\5{\hspace*{5mm}}
\def\2{{\textstyle\frac12}}
\def\4{{\textstyle\frac14}}

\begin{document}

\title{Edge States of a Periodic Chain with Four-Band Energy Spectrum}

\author{M. Eliashvili$^{1,2}$, D. Kereselidze$^1$, G. Tsitsishvili$^{1,2}$ and M. Tsitsishvili$^1$}
\affiliation{
$^{1}$Department of Physics, Tbilisi State University, Chavchavadze Ave. 3, Tbilisi 0179, Georgia\\
$^{2}$Razmadze Mathematical Institute, Tbilisi State University, Tamarashvili Str. 6, Tbilisi 0177, Georgia}

\begin{abstract}
Tight-binding model on a finite chain is studied with four-fold alternated hopping parameters $t_{1,2,3,4}$.
Imposing the open boundary conditions, the corresponding recursion is solved analytically with special
attention paid to the occurrence of edge states. Corresponding results are strongly corroborated by numeric
calculations. It is shown that in the system there exist four different edge phases if the number of sites is odd,
and eight edges phases if the chain comprises even number of sites. Phases are labelled by $\sigma_1\equiv{\rm sgn}(t_1t_3-t_2t_4)$,
$\sigma_2\equiv{\rm sgn}(t_1t_4-t_2t_3)$ and $\sigma_3\equiv{\rm sgn}(t_1t_2-t_3t_4)$. It is shown that these
quantities represent gauge invariant topological indices emerging in the corresponding infinite chains.
\end{abstract}

\maketitle

\section*{1. Introduction}

The most significant imprint of the topological order exhibited by topological insulators \cite{topins,shen,asboth} is the
occurrence of edge states. The simplest construction producing topological phases is the Su-Schrieffer-Heeger (SSH)
model originally introduced for describing the polyacetylene \cite{SSH}, and later on revisited in connection with topological
insulators. The SSH model exploits 1D tight-binding Hamiltonian with doubly alternated hopping parameters $t_{1,2}$ between
the nearest neighboring sites and produces two energy bands separated by a gap $2|t_1-t_2|$. Correspondingly, the two
topologically distinct phases occur, and the transition between those two takes
place at $t_1=t_2$ accompanied by closing and reopening the gap.

In the present paper we generalize the SSH model with the aim of detecting more diverse topological content
than the two-phase one. Similar attempt by involving the next-to-nearest hoppings has been carried out in Ref. [5]
leading to the phase diagram identical to the one of Haldane model \cite{H}. We study alternative modification
keeping the nearest neighbour hoppings only, but with four-fold alternated amplitudes $t_{1,2,3,4}$.
The corresponding energy spectrum consists of four bands separated by independently controlled gaps.
Since the closing and reopening of a gap is associated with topological phase transitions, one may expect the
occurrence of more rich topological content. In this scope we study a chain comprising finite number $N$ of sites,
and find 4 and 8 different phases for $N=odd$ and $N=even$, respectively.

The 8 phases (for $N=even$) are labelled by $(\sigma_1,\sigma_2,\sigma_3)$ where
\begin{align}
\sigma_1&\equiv{\rm sgn}(t_1t_3-t_2t_4),\nn\\
\sigma_2&\equiv{\rm sgn}(t_1t_4-t_2t_3),\nn\\
\sigma_3&\equiv{\rm sgn}(t_1t_2-t_3t_4),\nn
\end{align}
which are gauge invariant topological indices emerging in the
corresponding infinite chains. The 4 phases emerging in the case of $N=odd$ are labelled by $(\sigma_1,\sigma_2)$.

In Sec. 2 we present finite SSH model for demonstrating the technique \cite{BGL} of solving periodic three-term
recurrence relation analytically in terms of Chebyshev polynomials. In Sec. 3 the technique is applied to the four-band
model. Analytic results are corroborated by numeric calculations. In Sec. 4 we introduce certain scheme for constructing
Berry curvature for 1D periodic infinite chains. In the case of SSH (two-band) model the scheme gives out the Zak phase
\cite{Zak}, while in four-band model reconstructs the aforementioned labels $\sigma_{1,2,3}$ as gauge invariant
topological indices. Results are summarized in Sec. 5. Calculational details are placed in Appendix.

Chebyshev $U$-polynomials can be applied to 2D periodic lattices as well \cite{EJTT}.
Alternative analytic approach to the issue of edge modes can be found in Ref. [10].

\section*{2. Two-Band Finite Chain}

Finite chain with doubly alternated hoppings between the nearest sites is set by the tight-binding Hamiltonian
\be
H=-\sum_{n=1}^{N-1}t_n(c^\dag\hspace*{-1.25mm}_{n+1}c_n+c^\dag\hspace*{-1.25mm}_n c_{n+1})
\ee
where $N$ is the number of sites. Hopping parameters are periodic $t_{n+2}=t_n$ and can be put all positive.

The corresponding one-particle problem reads
\be
\left\lgroup\hspace*{-1mm}\ba{cccccc}
\epsilon & t_1 & 0 & 0 & \cdots & 0
\\
t_1 & \epsilon & t_2 & 0 & \cdots & 0
\\
0 & t_2 & \epsilon & t_1 & \cdots & 0
\\
0 & 0 & t_1 & \epsilon & \cdots & 0
\vspace*{-1.5mm}\\
\vdots & \vdots & \vdots & \vdots & & \vdots
\\
0 & 0 & 0 & 0 & \cdots & \epsilon
\ea\hspace*{-1mm}\right\rgroup
\hspace*{-1.5mm}\left\lgroup\hspace*{-1.5mm}\ba{c}
\psi_1
\\
\psi_2
\\
\psi_3
\\
\psi_4
\vspace*{-1.5mm}\\
\vdots
\\
\psi_N
\ea\hspace*{-1.5mm}\right\rgroup
=0.
\ee

In the component form this system appears as
\be
t_n\psi_n+\epsilon\psi_{n+1}+t_{n+1}\psi_{n+2}=0
\ee
supplied by the boundary conditions $\psi_0=\psi_{N+1}=0$.

Solution to (3) appears as (see Appendix A for details)
\babc
\begin{align}
\psi_{2n+1}&=\big[U_n(\xi)+(t_2/t_1)U_{n-1}(\xi)\big]\psi_1\\
\psi_{2n+2}&=-(\epsilon/t_1)U_n(\xi)\h\psi_1
\end{align}
\eabc
where
\be
\xi(t_1,t_2)=(\epsilon^2-t_1^2-t_2^2)/2t_1t_2
\ee
and $U_n$ are the Chebyshev polynomials of the second kind.

In (4) all $\psi_n$ are expressed via $\psi_1$. This is appropriate for describing the edge states localized at
left edge ($n=1$). For the states localized at the right edge ($n=N$) it is more convenient to express $\psi_n$
via $\psi_N$. For this purpose
we introduce $\psi_n=\phi_{N-n+1}$ and $t_n=u_{N-n}$. Then the
recurrence (3) takes the form
\be
u_n\phi_n+\epsilon\phi_{n+1}+u_{n+1}\phi_{n+2}=0
\ee
accompanied by $\phi_0=\phi_{N+1}=0$, and the solutions to (6) can be obtained by performing replacements
$t_n\to u_n$ and $\psi_n\to\phi_n$ in (4) and (5). We have $u_{1,2}=t_{1,2}$ for $N=even$, and $u_{1,2}=t_{2,1}$
for $N=odd$. Irrespectively of these options one finds $u_1^2+u_2^2=t_1^2+t_2^2$ and $u_1u_2=t_1t_2$,
where from we have $\xi(u_1,u_2)=\xi(t_1,t_2)$, and the solution to (6) appears as
\babc
\begin{align}
\phi_{2n+1}&=\big[U_n(\xi)+(u_2/u_1)U_{n-1}(\xi)\big]\phi_1,\\
\phi_{2n+2}&=-(\epsilon/u_1)U_n(\xi)\h\phi_1.
\end{align}
\eabc

Rewriting these in terms of $\phi_n=\psi_{N-n+1}$ we come to
\babc
\begin{align}
\psi_{N-2n}&=\big[U_n(\xi)+(u_2/u_1)U_{n-1}(\xi)\big]\psi_N,\\
\psi_{N-2n-1}&=-(\epsilon/u_1)U_n(\xi)\h\psi_N,
\end{align}
\eabc
where the concrete connection between $u_n$ and $t_n$ depends
on the number of sites, but $\xi$ is the same as given by (5).

Bulk-edge properties of eigenstates can be specified by the value of $\xi$.
For $|\xi|\leqslant1$ we put $\xi=\cos\gamma$ and using
\be
U_n(\cos\gamma)=\frac{\sin[(n+1)\gamma]}{\sin\gamma}
\ee
find that $\psi_n$ oscillates with respect to $n$, hence it is a bulk state.

For $|\xi|\geqslant1$ we put $\xi=\pm\cosh z$ and use
\be
U_n(\pm\cosh z)=(\pm1)^n\frac{\sinh[(n+1)z]}{\sinh z}.
\ee
Here we may have edge states due to the factors of $e^{\pm nz}$.

Taking $n=-1$ in (4b), the boundary condition $\psi_0=0$ is automatically satisfied due to $U_{-1}(x)=0$,
while the other one $\psi_{N+1}=0$ takes the role of secular equation and determines the energy spectrum.
Depending on the value of $N$, it appears as ($J$ is an integer)
\babc
\begin{align}
&N=2J+1:\5\epsilon\hh U_J(\xi)=0,\\
&N=2J+2:\5U_{J+1}(\xi)+(t_2/t_1)\hh U_J(\xi)=0.
\end{align}
\eabc

Similarly, taking $n=-1$ in (8b) we find that $\psi_{N+1}=0$ is automatically satisfied and the other boundary condition
takes the role of secular equation leading to the same (11).

\subsection*{ \it 2.1. Edge States for $N=2J+1$}

Secular equation (11a) breaks into two equations $\epsilon=0$ and $U_J(\xi)=0$.
Solutions to the later gives $|\xi|<1$ since the roots of $U_n(x)$ are all located in the
interval $(0,1)$. Hence the states determined by $U_J(\xi)=0$ are bulk states and
we turn to $\epsilon=0$.

Taking $\epsilon=0$ in (5) we find $\xi=-\frac12(t_1/t_2)-\frac12(t_2/t_1)<-1$, {\it i.e.}
these are edge states in accord with (10).

For $t_1<t_2$ we put $\xi=-\cosh z$ and resolve as $e^{-z}=t_1/t_2$. Using this in (4) we come to
\babc
\begin{align}
\psi_{2n+1}&=(-t_1/t_2)^n\psi_1,\\
\psi_{2n+2}&=0,
\end{align}
\eabc
which is localized at the left edge.

For $t_1>t_2$ we put $\xi=-\cosh z$ and resolve as $e^{-z}=t_2/t_1$. Using in (8) we come to
\babc
\begin{align}
\psi_{2n+1}&=(-t_2/t_1)^{J-n}\psi_N,\\
\psi_{2n+2}&=0,
\end{align}
\eabc
which is localized at the right edge.

Occurrence of a zero mode for $N=odd$ is a general fact: amount of zero modes in bipartite chains is
$|N_1-N_2|$ with $N_{1,2}$ the numbers of sites in two sublattices \cite{Lieb}.

\subsection*{\it 2.2. Edge States for $N=2J+2$}

We search for the edge states provided $J\to\infty$. For this purposes we study the equation (11b).
As already pointed out, the edge states emerge only in the cases when $|\xi|>1$.

\ding{70} $\xi>1$. In this case we put $\xi=\cosh z$ and rewrite (11b) as
\be
\frac{\sinh[(J+1)z]}{\sinh[Jz]}+\frac{t_2}{t_1}=0.
\ee
Assuming $z>0$, and taking the limit $J\to\infty$ we obtain
\be
e^{-z}+(t_1/t_2)=0
\ee
with no solution since $t_{1,2}>0$ (pointed out earlier),
hence $\xi>1$ cannot be realized.

\ding{70} $\xi<-1$. We then put $\xi=-\cosh z$ and rewrite (11b) as
\be
\frac{\sinh[(J+1)z]}{\sinh[Jz]}-\frac{t_2}{t_1}=0.
\ee
Assuming $z>0$, and taking the limit $J\to\infty$ we obtain
\be
e^{-z}=t_1/t_2
\ee
which (due to $z>0$) may be realized only for $t_1<t_2$.

Combining (17) with (5) we come to $\epsilon^2=0$.
The fact that the zero-mode occurs as $\epsilon^2=0$ signifies the energy level is doubly
degenerated. One wave function is obtained by use of (4) and appears as
\babc
\begin{align}
\psi_{2n+1}&=(-t_1/t_2)^n\psi_1,\\
\psi_{2n+2}&=0,
\end{align}
\eabc
representing the left edge state (localized at $\psi_1$).

The other is obtained by use of (8) and gives
\babc
\begin{align}
\psi_{2n+2}&=(-t_1/t_2)^{J-n}\psi_N,\\
\psi_{2n+1}&=0,
\end{align}
\eabc
representing the right edge state (localized at $\psi_N$).

\subsection*{\it 2.3. Numeric Calculations}

In support of the above analytic expressions below we bring the results of numeric calculations
for $N=101$ and $N=102$. The overall magnitude of $t_{1,2}$ is irrelevant and we
parameterize as $t_1=\sin\vartheta$ and $t_2=\cos\vartheta$. The energy spectra versus $\vartheta$ set by (2) are shown in Fig. 1.

\begin{center}
\includegraphics{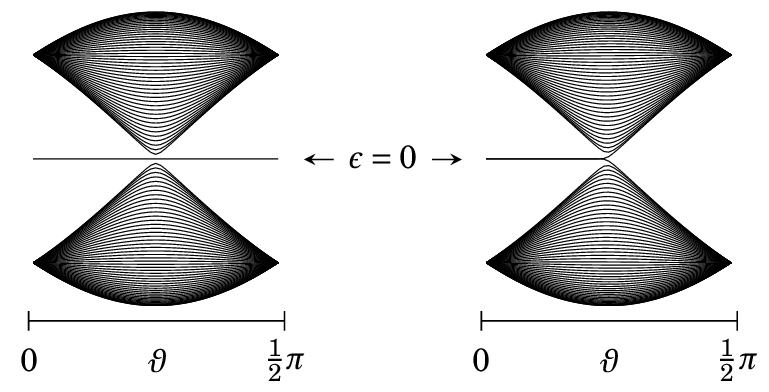}
\end{center}
\noindent
{\small FIG. 1: Energy spectra $\{\epsilon_1,\ldots,\epsilon_{101}\}$ (left) and
$\{\epsilon_1,\ldots,\epsilon_{102}\}$ (right) vs $\vartheta$ obtained
by numeric calculations for $N=101$ and $N=102$, respectively.
For $N=odd$ the single zero mode develops for any $t_1$ and $t_2$.
For $N=even$ doubly degenerated zero mode develops only for $t_1<t_2$.}

\section*{3. Four-Band Finite Chain}

We pass to a finite chain with four-fold alternated hoppings.
The corresponding one-particle equation reads as
\be
\left\lgroup\hspace*{-1mm}\ba{cccccccc}
\epsilon & t_1 & 0 & 0 & 0 & 0 & \cdots & 0
\\
t_1 & \epsilon & t_2 & 0 & 0 & 0 & \cdots & 0
\\
0 & t_2 & \epsilon & t_3 & 0 & 0 & \cdots & 0
\\
0 & 0 & t_3 & \epsilon & t_4 & 0 & \cdots & 0
\\
0 & 0 & 0 & t_4 & \epsilon & t_1 & \cdots & 0
\\
0 & 0 & 0 & 0 & t_1 & \epsilon & \cdots & 0
\vspace*{-1.5mm}\\
\vdots & \vdots & \vdots & \vdots & \vdots & \vdots & & \vdots
\\
0 & 0 & 0 & 0 & 0 & 0 & \cdots & \epsilon
\ea\hspace*{-1mm}\right\rgroup
\hspace*{-1.5mm}\left\lgroup\hspace*{-1.5mm}\ba{c}
\psi_1
\\
\psi_2
\\
\psi_3
\\
\psi_4
\\
\psi_5
\\
\psi_6
\vspace*{-1.5mm}\\
\vdots
\\
\psi_N
\ea\hspace*{-1.5mm}\right\rgroup
=0.
\ee

In the component form this system appears as
\be
t_n\psi_n+\epsilon\psi_{n+1}+t_{n+1}\psi_{n+2}=0
\ee
with $t_{n+4}=t_n$, and are supplied by $\psi_0=\psi_{N+1}=0$.

In accord with Ref. [7] we write the solution to (21)
\babc
\begin{align}
\psi_{4n+1}&=\Bigg[U_n(\xi)+\frac{t_4}{t_3}\hh\frac{\epsilon^2-t_2^2}{t_1t_2}\hh U_{n-1}(\xi)\Bigg]\psi_1,\\
\psi_{4n+2}&=-\frac{\epsilon}{t_1}\Bigg[U_n(\xi)+\frac{t_1t_4}{t_2t_3}\hh U_{n-1}(\xi)\Bigg]\psi_1,\\
\psi_{4n+3}&=\Bigg[\frac{\epsilon^2-t_1^2}{t_1t_2}\hh U_n(\xi)+\frac{t_4}{t_3}\hh U_{n-1}(\xi)\Bigg]\psi_1,\\
\psi_{4n+4}&=-\epsilon\hh\frac{\epsilon^2-t_1^2-t_2^2}{t_1t_2t_3}\hh U_n(\xi)\hh\psi_1,
\end{align}
\eabc
where
\be
\xi\equiv\frac{\epsilon^4-(t_1^2+t_2^2+t_3^2+t_4^2)\epsilon^2+t_1^2t_3^2+t_2^2t_4^2}{2t_1t_2t_3t_4}.
\ee

Alternatively, we can express $\psi_n$ in terms of $\psi_N$. Introducing $\psi_n=\phi_{N-n+1}$ and $t_n=u_{N-n}$
we rewrite (21) as
\be
u_n\phi_n+\epsilon\phi_{n+1}+u_{n+1}\phi_{n+2}=0
\ee
accompanied by $\phi_0=\phi_{N+1}=0$.

Solution to (24) can be written by replacing $t_n\to u_n$ and $\psi_n\to\phi_n$ in (22) and (23).
The parameter $\xi$ defined by (23) is invariant with respect to $t_n\to u_n$, and we come to
\babc
\begin{align}
\psi_{N-4n}&=\Bigg[U_n(\xi)+\frac{u_4}{u_3}\hh\frac{\epsilon^2-u_2^2}{u_1u_2}\hh U_{n-1}(\xi)\Bigg]\psi_N,\\
\psi_{N-4n-1}&=-\frac{\epsilon}{u_1}\Bigg[U_n(\xi)+\frac{u_1u_4}{u_2u_3}\hh U_{n-1}(\xi)\Bigg]\psi_N,\\
\psi_{N-4n-2}&=\Bigg[\frac{\epsilon^2-u_1^2}{u_1u_2}\hh U_n(\xi)+\frac{u_4}{u_3}\hh U_{n-1}(\xi)\Bigg]\psi_N,\\
\psi_{N-4n-3}&=-\epsilon\hh\frac{\epsilon^2-u_1^2-u_2^2}{u_1u_2u_3}\hh U_n(\xi)\hh\psi_N,
\end{align}
\eabc
where the connection between $u_n$ and $t_n$ depends on $N$.

Taking $n=-1$ in (22d) the condition $\psi_0=0$ is automatically satisfied due to $U_{-1}(x)=0$,
while the one $\psi_{N+1}=0$ appears as
\babc
\begin{align}
&N=4J+1:\hspace*{3mm}\epsilon\big[t_2t_3U_J(\xi)+t_1t_4U_{J-1}(\xi)\big]=0,\\
&N=4J+2:\hspace*{3mm}t_3(\epsilon^2-t_1^2)U_J(\xi)+t_1t_2t_4U_{J-1}(\xi)=0,\\
&N=4J+3:\hspace*{3mm}\epsilon(\epsilon^2-t_1^2-t_2^2)U_J(\xi)=0,\\
&N=4J+4:\hspace*{3mm}t_1t_2 t_3U_{J+1}(\xi)+t_4(\epsilon^2-t_2^2)U_J(\xi)=0,
\end{align}
\eabc
and determines the energy eigenvalues.

Taking $n=-1$ in (25d) we find $\psi_{N+1}=0$ is automatically satisfied, while $\psi_0=0$
leads to the same (26).

In the following four subsections we separately present the cases $N=4J+j$ with $j=1,2,3,4$.
In order expressions to be compact, hereafter we put $\psi_1=1$ and $\psi_N=1$ for the left and right edge states respectively,
and indicate only non-vanishing components of $\psi_n$.

\subsection*{\it 3.1. Edge States for $N=4J+1$}

Calculational details are collected in Appendix B, while here we show the results.

For $t_1t_3<t_2t_4$ one left edge state occurs
\be
\epsilon=0,\hspace*{2.5mm}\left\{\hspace*{-1mm}
\ba{l}\psi_{4n+1}=(t_1t_3/t_2t_4)^n,\\\psi_{4n+3}=-(t_1/t_2)(t_1t_3/t_2t_4)^n.\ea\right.
\ee

For $t_1t_3>t_2t_4$ we have one right edge state
\be
\epsilon=0,\hspace*{2.5mm}\left\{\hspace*{-1mm}
\ba{l}\psi_{4n+1}=(t_2t_4/t_1t_3)^{J-n},\\\psi_{4n+4}=-(t_4/t_3)(t_2t_4/t_1t_3)^{J-n-1}.\ea\right.
\ee

For $t_1t_4>t_2t_3$ we have four edge states. Two states out of those four are localized at $n=1$ and appear as
\babc
\be
\hspace*{-2mm}\epsilon=\pm(t_1^2+t_2^2)^{1/2},\hspace*{2.5mm}
\left\{\hspace*{-1mm}\ba{l}
\psi_{4n+1}=(-t_2t_3/t_1t_4)^n,
\\
\psi_{4n+2}=-(\epsilon/t_1)(-t_2t_3/t_1t_4)^n,
\\
\psi_{4n+3}=(t_2/t_1)(-t_2t_3/t_1t_4)^n.
\ea\right.
\ee
The other two are localized at $n=N$ and look as
\be
\hspace*{-2mm}\epsilon=\pm(t_3^2+t_4^2)^{1/2},\hspace*{2.5mm}
\left\{\hspace*{-1mm}\ba{l}
\psi_{4n+1}=(-t_2t_3/t_1t_4)^{J-n},
\\
\psi_{4n+3}=(t_3/t_4)(-t_2t_3/t_1t_4)^{J-n-1},
\\
\psi_{4n}=-(\epsilon/t_4)(-t_2t_3/t_1t_4)^{J-n}.
\ea\right.
\ee
\eabc

Introduce $\sigma_1\equiv{\rm sgn}(t_1t_3-t_2t_4)$ and $\sigma_2\equiv{\rm sgn}(t_1t_4-t_2t_3)$.
We then label the $4$ phases by $(\sigma_1,\sigma_2)$ as follows
\be\ba{lll}
A&(-,-):&\textrm{1 \it edge state (27)},\\
B&(-,+):&\textrm{5 \it edge states (27), (29)},\\
C&(+,-):&\textrm{1 \it edge state (28)},\\
D&(+,+):&\textrm{5 \it edge states (28), (29)}.
\ea\nn\ee

In order to confirm the analytic expressions, we present the numerically obtained energy spectra
for $N=101$. We parameterize $t_{1,2,3,4}$ as (overall magnitude irrelevant)
\babc
\begin{align}
t_1&=\cos\vartheta_3\sin\vartheta_1,
\\
t_2&=\cos\vartheta_3\cos\vartheta_1,
\\
t_3&=\sin\vartheta_3\cos\vartheta_2,
\\
t_4&=\sin\vartheta_3\sin\vartheta_2.
\end{align}
\eabc

In Fig. 2 we present the energy spectrum $\epsilon_{1\div101}$ versus $\vartheta_1$
with $\vartheta_2=\frac18\pi$ and $\vartheta_3=\frac{1}{16}\pi$. For $\vartheta_1<\vartheta_2$
the system is found in phase $A$. Increasing $\vartheta_1$ we observe the transition into phase $C$ at
$\vartheta_1=\vartheta_2$, and subsequently into phase $D$ at $\vartheta_1=\frac12\pi-\vartheta_2$.
Thus the width of phase $C$ laying in between the phases $A$ and $D$ is $\frac12\pi-2\vartheta_2$.
\begin{center}
\includegraphics{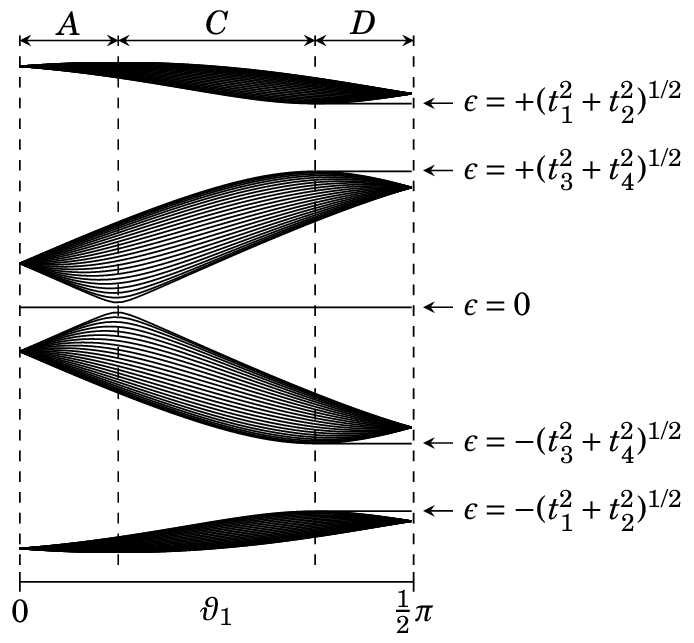}
\end{center}
\noindent
{\small FIG. 2: Energy spectrum $\epsilon_{1\div101}$ vs $\vartheta_1$
for $\vartheta_2=\frac18\pi$ and $\vartheta_3=\frac{3}{16}\pi$.}
\vspace*{3mm}

We comment on how the dispersion shown in Fig. 2 is affected by varying $\vartheta_2$.
Increasing $\vartheta_2$ the phase $C$ shrinks and disappears at $\vartheta_2=\frac14\pi$.
In that case the system passes from phase $A$ right into $D$. Further increase of $\vartheta_2$
cause the phases $A$ and $D$ become separated again, but with phase $B$ in between as shown in Fig. 3.
\begin{center}
\includegraphics{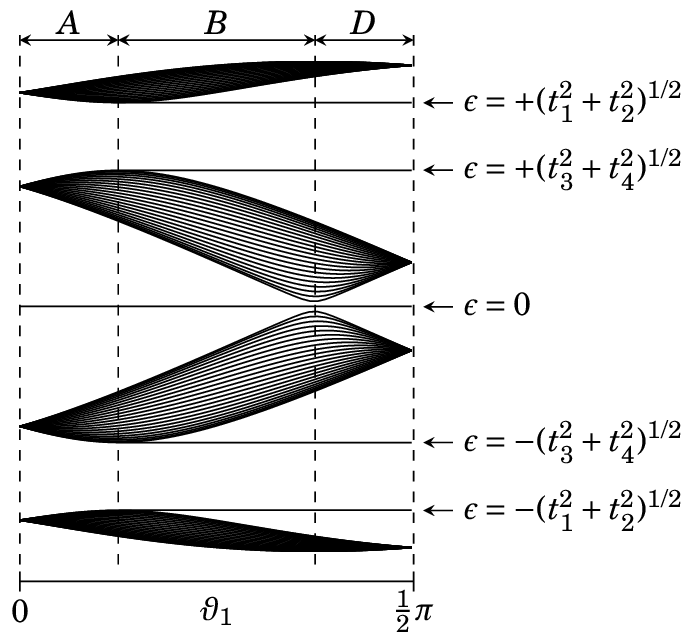}
\end{center}
\noindent
{\small FIG. 3: Energy spectrum $\epsilon_{1\div101}$ vs $\vartheta_1$
for $\vartheta_2=\frac38\pi$ and $\vartheta_3=\frac{3}{16}\pi$.}
\vspace*{3mm}

We discuss the role of $\vartheta_3$. In Fig. 2 and Fig. 3 there are gaps due to
$(t_1^2+t_2^2)^{1/2}>(t_3^2+t_4^2)^{1/2}$. The later results from
$\vartheta_3<\frac14\pi$ (see (30)). The gaps close at $\vartheta_3=\frac14\pi$,
as depicted in Fig. 4 with $\vartheta_2=\frac18\pi$ (top) and $\frac38\pi$ (bottom).
Behaviour with respect to $\vartheta_2$ is the same, as already stated.
\begin{center}
\includegraphics{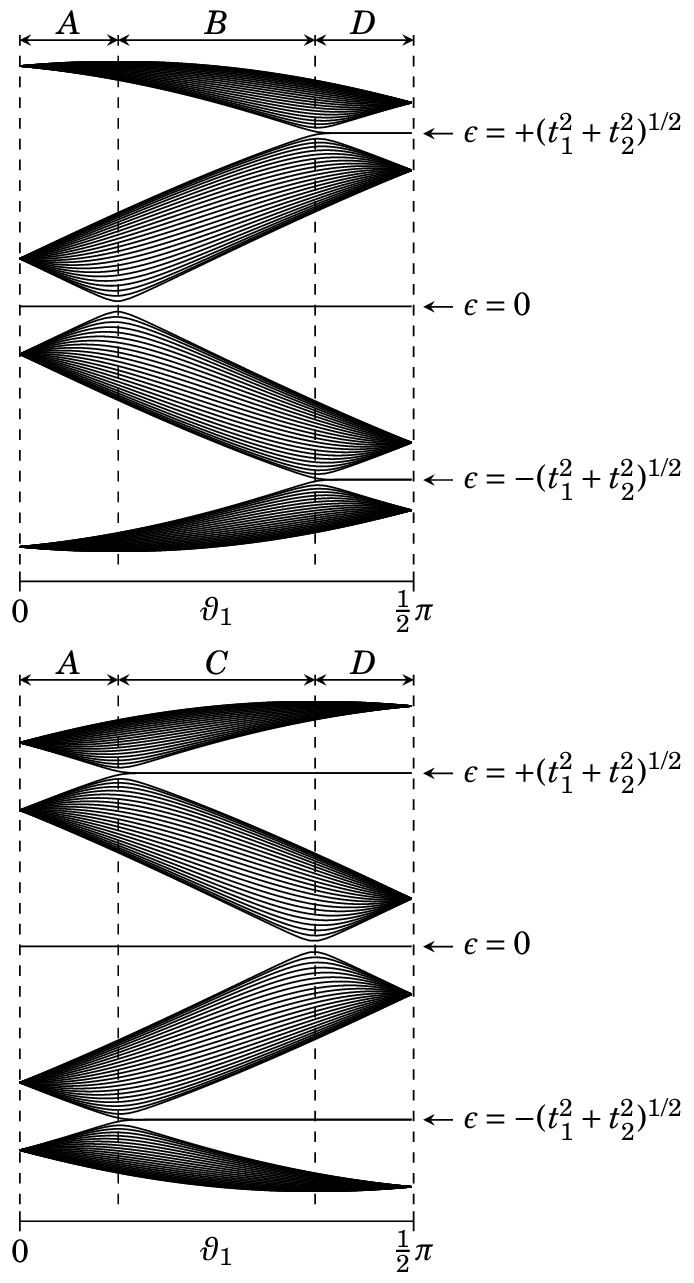}
\end{center}
\noindent
{\small FIG. 4: $\epsilon_{1\div101}$ vs $\vartheta_1$
for $\vartheta_3=\frac14\pi$. $\vartheta_2=\frac18\pi$ (top), $\vartheta_2=\frac38\pi$ (bottom).}

Further increase of $\vartheta_3$ result in reopening the gaps with the only difference that we have
$(t_3^2+t_4^2)^{1/2}>(t_1^2+t_2^2)^{1/2}$ due to $\vartheta_3>\frac14\pi$ (see (30)).

\subsection*{\it 3.2. Edge States for $N=4J+2$}

For $t_1t_3<t_2t_4$ we have two edge states (Appendix C)
\babc
\begin{align}
&\epsilon=0\h,\hspace*{2mm}
\left\{\hspace*{-1mm}\ba{l}
\psi_{4n+1}=(t_1t_3/t_2t_4)^n,
\\
\psi_{4n+3}=-(t_1/t_2)(t_1t_3/t_2t_4)^n,
\ea\right.
\\
&\epsilon=0\h,\hspace*{2mm}
\left\{\hspace*{-1mm}\ba{l}
\psi_{4n+2}=(t_1t_3/t_2t_4)^{J-n},
\\
\psi_{4n+4}=-(t_1/t_4)(t_1t_3/t_2t_4)^{J-n-1}.
\ea\right.
\end{align}
\eabc
localized at left and right edges, respectively.

For $t_1t_4>t_2t_3$ we have two edge states
\be
\epsilon=\pm(t_1^2+t_2^2)^{1/2},\hspace*{2mm}
\left\{\hspace*{-1mm}\ba{l}
\psi_{4n+1}=(-t_2t_3/t_1t_4)^n,
\\
\psi_{4n+2}=(-\epsilon/t_1)(-t_2t_3/t_1t_4)^n,
\\
\psi_{4n+3}=(t_2/t_1)(-t_2t_3/t_1t_4)^n.
\ea\right.
\ee

For $t_1t_2>t_3t_4$ we have two edge states
\be
\epsilon=\pm(t_1^2+t_4^2)^{1/2},\hspace*{2mm}
\left\{\hspace*{-1mm}\ba{l}
\psi_{4n+2}=(-t_3t_4/t_1t_2)^{J-n},
\\
\psi_{4n+1}=(-\epsilon/t_1)(-t_3t_4/t_1t_2)^{J-n},
\\
\psi_{4n+4}=(t_4/t_1)(-t_3t_4/t_1t_2)^{J-n-1}.
\ea\right.
\ee

Here we employ three parameters $\sigma_1\equiv{\rm sgn}(t_1t_3-t_2t_4)$, $\sigma_2\equiv{\rm sgn}(t_1t_4-t_2t_3)$,
$\sigma_3\equiv{\rm sgn}(t_1t_2-t_3t_4)$ and label the eight ($=2^3$) phases by $(\sigma_1,\sigma_2,\sigma_3)$
as follows
\be\ba{lll}
\textrm{\ding{172}}&(-,-,-):&\textrm{2 \it edge states (31)},\\
\textrm{\ding{173}}&(-,-,+):&\textrm{4 \it edge states (31), (33)},\\
\textrm{\ding{174}}&(-,+,-):&\textrm{4 \it edge states (31), (32)},\\
\textrm{\ding{175}}&(-,+,+):&\textrm{6 \it edge states (31), (32), (33)},\\
\textrm{\ding{176}}&(+,-,-):&\textrm{\it no edge states},\\
\textrm{\ding{177}}&(+,-,+):&\textrm{2 \it edge states (33)},\\
\textrm{\ding{178}}&(+,+,-):&\textrm{2 \it edge states (32)},\\
\textrm{\ding{179}}&(+,+,+):&\textrm{4 \it edge states (32), (33)}.
\ea\nn\ee

Fig. 5 depicts the spectrum versus $\vartheta_1$ with $\vartheta_2=\frac18\pi$ and $\vartheta_3=\frac{3}{16}\pi$.
\begin{center}
\includegraphics{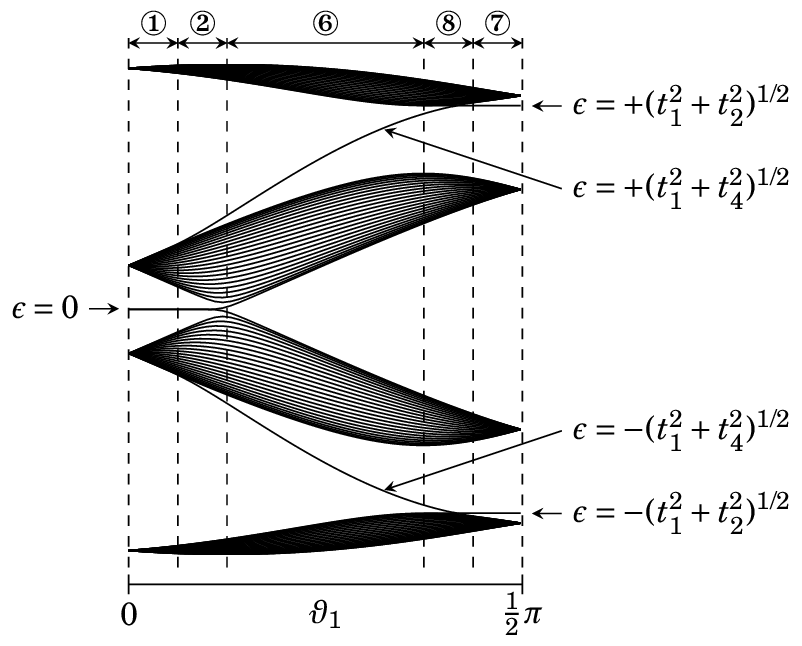}
\end{center}
\noindent
{\small FIG. 5: Energy spectrum $\epsilon_{1\div102}$ vs $\vartheta_1$ for $\vartheta_2=\frac18\pi$ and $\vartheta_3=\frac{3}{16}\pi$.}

\newpage

Increasing $\vartheta_2$, the width of phase \ding{178} ($\frac12\pi-2\vartheta_2$) shrinks and disappears at
$\vartheta_2=\frac14\pi$, so that the phase \ding{174} becomes followed right by \ding{179}. Further increase of
$\vartheta_2$ causes the phases \ding{174} and \ding{179} become detached again, but with \ding{175} instead
of \ding{178} in between as shown in Fig. 6.
\begin{center}
\includegraphics{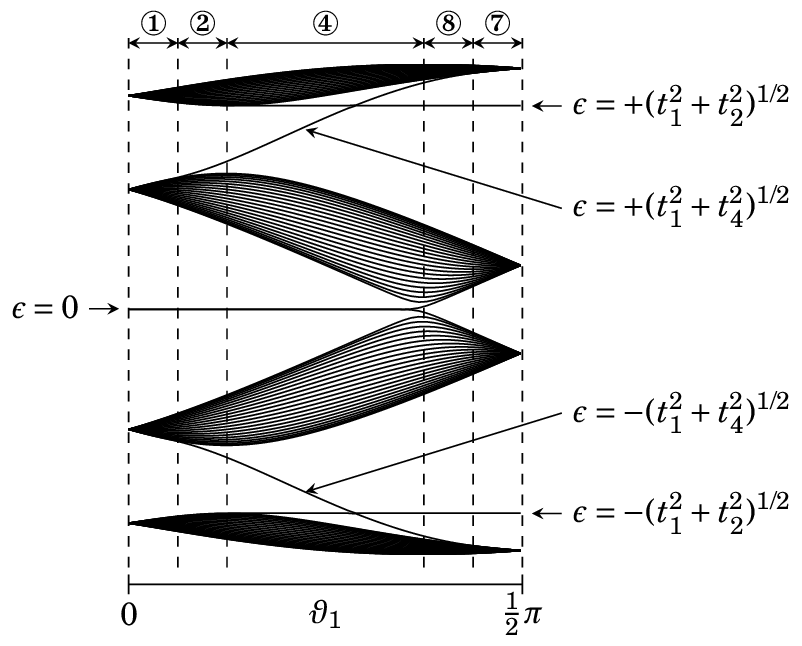}
\end{center}
\noindent
{\small FIG. 6: Energy spectrum $\epsilon_{1\div102}$ vs $\vartheta_1$ for $\vartheta_2=\frac38\pi$ and $\vartheta_3=\frac{3}{16}\pi$.}
\vspace*{3mm}

Dependence on $\vartheta_3$ is analogous to the case of $N=4J+1$.
Namely, for $\vartheta_3<\frac14\pi$ (Fig. 5 and Fig. 6) the two bunches of positive (negative)
levels are separated by the gap, but with the levels $\pm(t_1^2+t_4^2)^{1/2}$ in between.
Increasing $\vartheta_3$ we find that for $\vartheta_3=\frac14\pi$ the gaps close
and for $\vartheta_3>\frac14\pi$ reopen as shown in Fig. 7,
but now the energy levels $\pm(t_1^2+t_4^2)^{1/2}$ are lost.

\begin{center}
\includegraphics{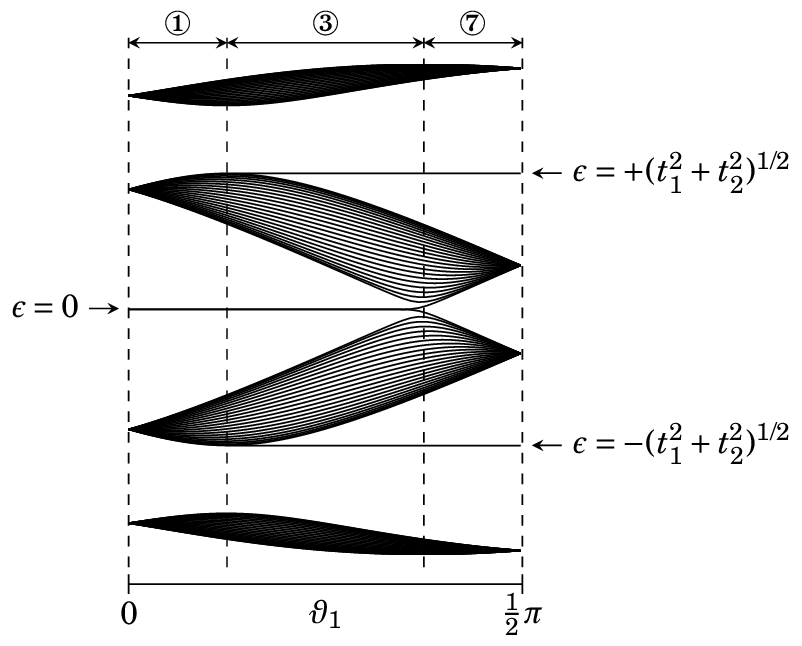}
\end{center}
\noindent
{\small FIG. 7: Energy spectrum $\epsilon_{1\div102}$ vs $\vartheta_1$
for $\vartheta_2=\frac38\pi$ and $\vartheta_3=\frac{5}{16}\pi$.}

\subsection*{\it 3.3. Edge States for $N=4J+3$}

For $t_1t_3<t_2t_4$ we have the left edge state (Appendix D)
\be
\epsilon=0\h,\hspace*{2mm}
\left\{\hspace*{-1mm}\ba{l}
\psi_{4n+1}=(t_1t_3/t_2t_4)^n,
\\
\psi_{4n+3}=(-t_1/t_2)(t_1t_3/t_2t_4)^n.
\ea\right.
\ee

For $t_1t_3>t_2t_4$ we have the right edge state
\be
\epsilon=0\h,\hspace*{2mm}
\left\{\hspace*{-1mm}\ba{l}
\psi_{4n+3}=(t_2t_4/t_1t_3)^{J-n},
\\
\psi_{4n+1}=(-t_2/t_1)(t_2t_4/t_1t_3)^{J-n}.
\ea\right.
\ee

For $t_1t_4>t_2t_3$ we have two left edge states
\be
\epsilon=\pm(t_1^2+t_2^2)^{1/2},\hspace*{2mm}
\left\{\hspace*{-1mm}\ba{l}
\psi_{4n+1}=(-t_2t_3/t_1t_4)^n,
\\
\psi_{4n+2}=(-\epsilon/t_1)(-t_2t_3/t_1t_4)^n,
\\
\psi_{4n+3}=(t_2/t_1)(-t_2t_3/t_1t_4)^n,
\ea\right.
\ee
and for $t_1t_4<t_2t_3$ we have two right edge states
\be
\epsilon=\pm(t_1^2+t_2^2)^{1/2},\hspace*{2mm}
\left\{\hspace*{-1mm}\ba{l}
\psi_{4n+3}=(-t_1t_4/t_2t_3)^{J-n},
\\
\psi_{4n+2}=(-\epsilon/t_2)(-t_1t_4/t_2t_3)^{J-n},
\\
\psi_{4n+1}=(t_1/t_2)(-t_1t_4/t_2t_3)^{J-n}.
\ea\right.
\ee

Using $\sigma_1={\rm sgn}(t_1t_3-t_2t_4)$ and  $\sigma_2={\rm sgn}(t_1t_4-t_2t_3)$ as in the case of $N=4J+1$,
we label the 4 phases by $A,B,C,D$ with respect to $(\sigma_1,\sigma_2)$.

Properties of spectra with respect to $\vartheta_2$ and $\vartheta_3$ are the same as
for $N=4J+1$. Therefore, in Fig. 8 we show only couple of cases
$(\vartheta_2,\vartheta_3)=(\frac18\pi,\frac{3}{16}\pi)$ and
$(\vartheta_2,\vartheta_3)=(\frac38\pi,\frac{5}{16}\pi)$.

\begin{center}
\includegraphics{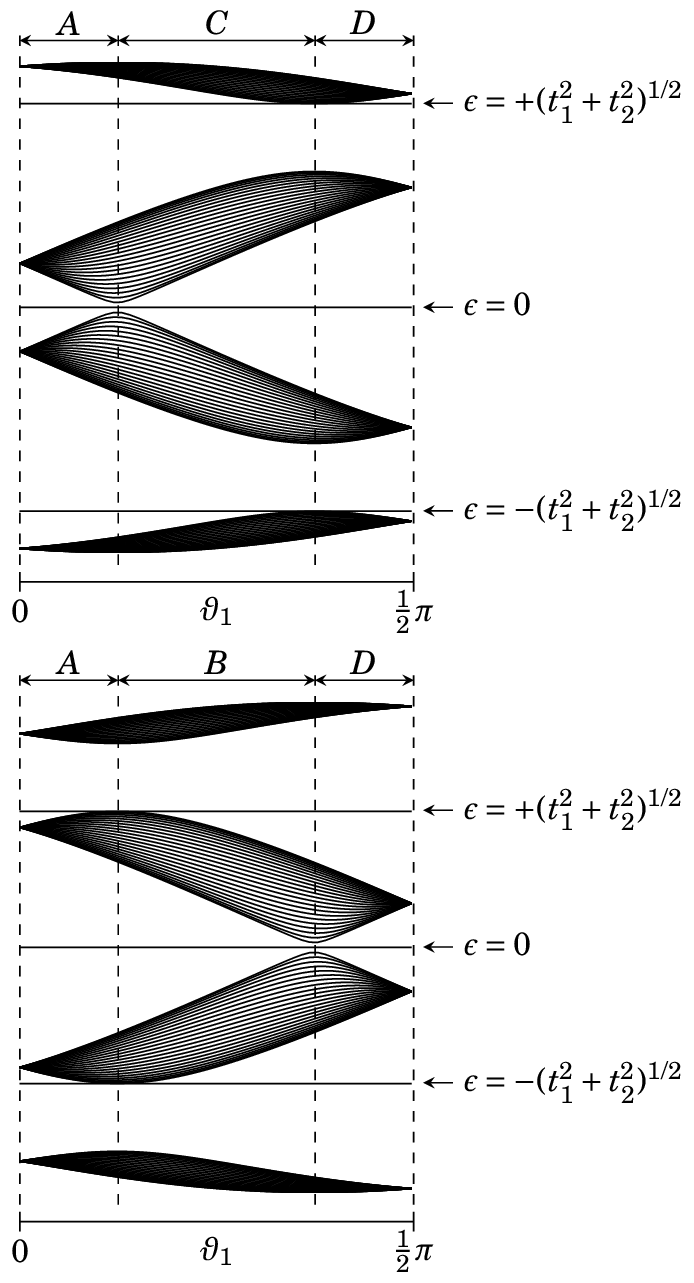}
\end{center}
\noindent
{\small FIG. 8: The spectrum $\epsilon_{1\div103}$ vs $\vartheta_1$
for $(\vartheta_2,\vartheta_3)=(\frac18\pi,\frac{3}{16}\pi)$ (top)
and $(\vartheta_2,\vartheta_3)=(\frac38\pi,\frac{5}{16}\pi)$ (bottom).}

In the top panel we have $\vartheta_2<\frac14\pi$ and consequently the phase $C$ occurs in between $A$ and $D$.
For $\vartheta_2=\frac14\pi$ the intermediate phase disappears and reappears as $B$ for $\vartheta_2>\frac14\pi$
in the bottom panel.

For $\vartheta_3<\frac14\pi$ (top) there are gaps, which close at $\vartheta_3=\frac14\pi$
and reopen for $\vartheta_3>\frac14\pi$ (bottom). After reopening the gaps the edge state levels
$\epsilon=\pm(t_1^2+t_2^2)^{1/2}$ initially attached to the outer bands (top pannel) become attached
to the inner ones (bottom panne).

\subsection*{\it 3.4. Edge States for $N=4J+4$}

For $t_1t_3<t_2t_4$ we have two edge states (see Appendix E). One is the left edge state and appears as
\babc
\be
\epsilon=0\h,\hspace*{2mm}
\left\{\hspace*{-1mm}\ba{l}
\psi_{4n+1}=(t_1t_3/t_2t_4)^n,
\\
\psi_{4n+3}=-(t_1/t_2)(t_1t_3/t_2t_4)^n.
\ea\right.
\ee
The other is the right edge state and looks as
\be
\epsilon=0\h,\hspace*{2mm}
\left\{\hspace*{-1mm}\ba{l}
\psi_{4n+4}=(t_1t_3/t_2t_4)^{J-n},
\\
\psi_{4n+2}=-(t_3/t_2)(t_1t_3/t_2t_4)^{J-n}.
\ea\right.
\ee
\eabc
For $t_1t_4>t_2t_3$ we have
\be
\epsilon=\pm(t_1^2+t_2^2)^{1/2},\hspace*{2mm}
\left\{\hspace*{-1mm}\ba{l}
\psi_{4n+1}=(-t_2t_3/t_1t_4)^n,
\\
\psi_{4n+2}=(-\epsilon/t_1)(-t_2t_3/t_1t_4)^n,
\\
\psi_{4n+3}=(t_2/t_1)(-t_2t_3/t_1t_4)^n.
\ea\right.
\ee
For $t_3t_4>t_1t_2$ we have
\be
\epsilon=\pm(t_2^2+t_3^2)^{1/2},\hspace*{2mm}
\left\{\hspace*{-1mm}\ba{l}
\psi_{4n+4}=(-t_1t_2/t_3t_4)^{J-n},
\\
\psi_{4n+3}=(-\epsilon/t_3)(-t_1t_2/t_3t_4)^{J-n},
\\
\psi_{4n+2}=(t_2/t_3)(-t_1t_2/t_3t_4)^{J-n}.
\ea\right.
\ee

Here we have the same defining parameters $(\sigma_1,\sigma_2,\sigma_3)$ as in the case of $N=4J+2$,
hence $8$ phases, labelled by \ding{172}$\div$\ding{179}
with respect to the values of $(\sigma_1,\sigma_2,\sigma_3)$ as for $N=4J+2$.
\begin{center}
\includegraphics{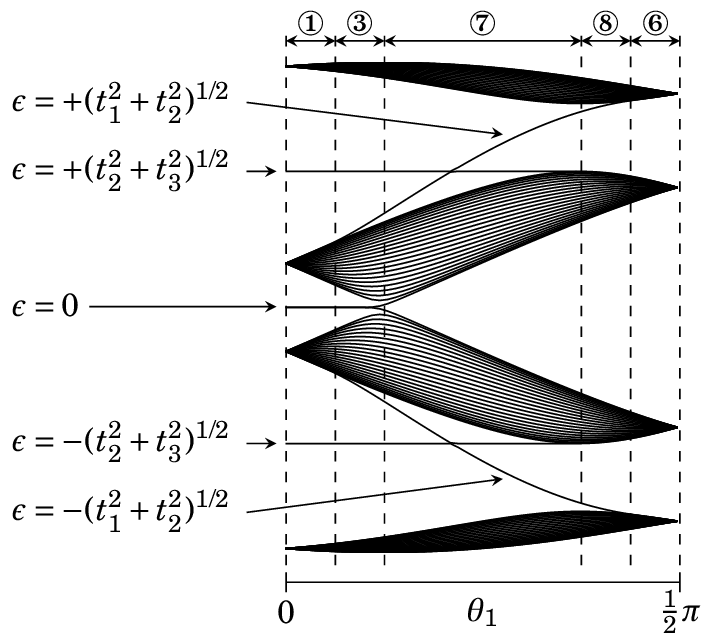}
\end{center}
\noindent
{\small FIG. 9: Energy spectrum $\epsilon_{1\div104}$ vs $\theta_1$
for $\theta_2=\frac18\pi$ and $\theta_3=\frac{3}{16}\pi$.}

Plotting the spectra here we use the parametrization
\babc
\begin{align}
t_1&=\cos\theta_3\sin\theta_1,
\\
t_2&=\sin\theta_3\sin\theta_2,
\\
t_3&=\sin\theta_3\cos\theta_2,
\\
t_4&=\cos\theta_3\cos\theta_1.
\end{align}
\eabc

The case of $N=4J+4$ is similar to $N=4J+2$, and we present the two cases only, just for the sake of presentation:
$(\theta_2,\theta_3)=(\frac18\pi,\frac{3}{16}\pi)$ in Fig. 9 and
$(\theta_2,\theta_3)=(\frac18\pi, \frac{5}{16}\pi)$ in Fig. 10.
\begin{center}
\includegraphics{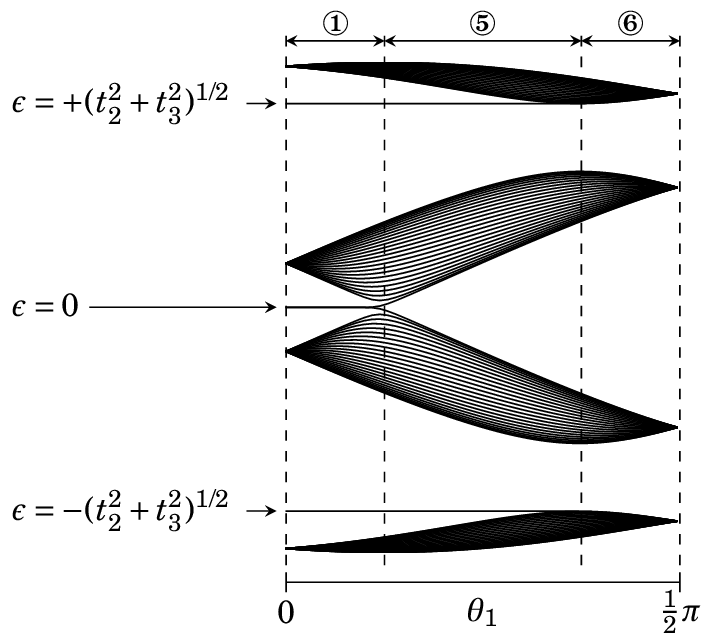}
\end{center}
\noindent
{\small FIG. 10: Energy spectrum $\epsilon_{1\div104}$ vs $\theta_1$
for $\theta_2=\frac18\pi$ and $\theta_3=\frac{5}{16}\pi$.}

\section*{4. Topological Invariants}

Edges phases are usually characterized by certain topological invariants. In 2D lattices these can be constructed
by introducing the Berry connections
\be
A_{1,2}=\frac{i}{\pi}\hh\psi^\dag\frac{\partial\psi}{\partial k_{1,2}}
\ee
where $k_{1,2}$ are the component of 2D quasi-momentum.
Then the gauge invariant Berry curvature is defined as
\be
F=\frac{\partial A_2}{\partial k_1}-\frac{\partial A_1}{\partial k_2}
\ee
which after integrating over the torus $k_{1,2}\in[0,2\pi]$ leads to gauge invariant topological indices.
Such construction is applicable irrespectively of the number of energy bands ({\it e.g.} [12] for 2D lattices with four-band
energy spectra).

The obstacle in constructing curvatures for 1D periodic systems is the lack of $k_2$. This drawback can be overcome by
introducing additional parameter $\alpha$ so the curvature is constructed in terms of $(k,\alpha)$ instead of $(k_1,k_2)$.
Such an approach is employed {\it e.g.} in Ref. [5]. For the SSH model the choice of extra parameter as $\alpha=t_1-t_2$
has been commented in lecture notes by E. Mele (unpublished). However, it is not clear how to generalize this choice to a model
comprising four hopping parameters  $t_{1,2,3,4}$. Besides, we would expect the parameter $\alpha$ to vary on a circle like
$k_1$ and $k_2$ do in the case of 2D.

We propose a general scheme of constructing gauge invariant topological indices for 1D periodic chains. We first
examine the scheme on two-band chain and reproduce the Zak phase. We then apply it to the four-band model under
consideration and show that the quantities $\sigma_{1,2,3}$ labeling the edge phases of the finite chain represent
gauge invariant topological indices.

\subsection*{\it 4.1. Two-Band Chain}

Consider the tight-binding Hamiltonian
\be
H=\sum_nt_n(c^\dag\hspace*{-1.3mm}_{n+1}c_n+h.c.)
\ee
where $t_n$ are periodic $t_{n+2}=t_n$ and can be presented as
\be
t_n=\2(t_1+t_2)-\2(t_1-t_2)\cos(\pi n).
\ee

Introduce the parameter $\alpha$ by generalizing $t_n$ as
\be
t_n\to \tau_n(\alpha)=\2(t_1+t_2)-\2(t_1-t_2)\cos(\pi n+\alpha).
\ee
For $\alpha=0$ we have $\tau_{1,2}=t_{1,2}$, and $\tau_{1,2}=t_{2,1}$ for $\alpha=\pi$
what represents the redefinition of a unit cell ($t_1\rightleftarrows t_2$).
Increasing $\alpha$ up to $2\pi$ we return back to $\tau_{1,2}=t_{1,2}$.
This construction is trivially extendable to any 1D periodic chain.

Rewriting (44) in the Fourier form, the corresponding one-particle Hamiltonian appears as
\be
\mathcal H=\left\lgroup\ba{cc}0&\tau_1e^{+ik\ell}+\tau_2e^{-ik\ell}
\\\vspace*{-3mm}\\
\tau_1e^{-ik\ell}+\tau_2e^{+ik\ell}&0\ea\right\rgroup.
\ee
where $\ell$ is the separation between the neighbouring sites, {\it i.e.} $2\ell$ is the period of the chain.
In what follows we use the dimensionless momentum $\kappa\equiv(2\ell)k$ with $-\pi\leqslant\kappa\leqslant+\pi$.

Eigenvalues and normalized eigenvectors are given by
\babc
\begin{align}
&\epsilon^2(\kappa,\alpha)=\tau_1^2+\tau_2^2+2\tau_1^{}\tau_2^{}\cos\kappa,\\
&\psi(\kappa,\alpha)=\frac{1}{\sqrt{2\epsilon^2}}\hspace*{-1mm}\left\lgroup\ba{c}\tau_1e^{+(i/2)\kappa}+\tau_2e^{-(i/2)\kappa}\\\vspace*{-3mm}\\\epsilon\ea\right\rgroup\hspace*{-0.5mm},
\end{align}
\eabc
hence the system is defined on a torus $\kappa\in[0,2\pi]$, $\alpha\in[0,2\pi]$ depicted in Fig. 11.
\begin{center}
\includegraphics{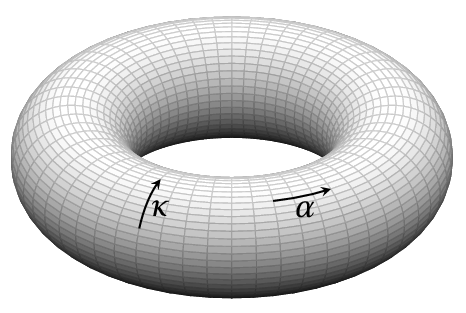}\\
{\small FIG. 11: Torus formed by $0\leqslant\kappa\leqslant2\pi$ and $0\leqslant\alpha\leqslant2\pi$.}
\end{center}
\vspace*{3mm}

Berry connections are defined as
\babc
\begin{align}
A_\kappa&=\frac{i}{\pi}\h\psi^\dag\frac{\partial\psi}{\partial\kappa}\\
A_\alpha&=\frac{i}{\pi}\h\psi^\dag\frac{\partial\psi}{\partial\alpha}
\end{align}
\eabc
and the gauge invariant curvature is constructed as
\be
F=\partial_\kappa A_\alpha-\partial_\alpha A_\kappa.
\ee

Let $\Omega$ be the surface on the torus bounded by two closed contours corresponding to some $\alpha_{1,2}$
as shown in Fig. 12.
\begin{center}
\includegraphics{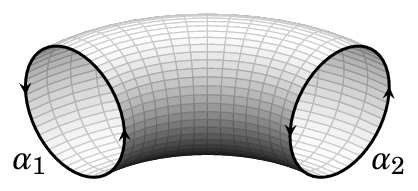}\\
{\small FIG. 12: The surface $\Omega$ bounded by two closed contours.}
\end{center}
\vspace*{3mm}

Integrating the curvature (50) over $\Omega$ and using Stokes' theorem we find
\be
\mu\equiv\int_\Omega Fd\kappa d\alpha=\oint A_\kappa(\kappa,\alpha_1)d\kappa-\oint A_\kappa(\kappa,\alpha_2)d\kappa
\ee
where the right hand side represents the gauge invariant quantity usually referred to as the Zak phase.

Employing (48) we find
\be
A_\kappa(\kappa,\alpha)=\frac{1}{4\pi}\cdot\frac{\tau_2^2-\tau_1^2}{\tau_1^2+\tau_2^2+2\tau_1^{}\tau_2^{}\cos\kappa}.
\ee
Using this in (51) with $\alpha_1=0$ and $\alpha_2=\pi$ we find
\be
\mu={\rm sgn}(t_2^2-t_1^2).
\ee
{\it i.e.} the Zak phase takes the values $\pm1$.

\subsection*{\it 4.2. Four-Band Chain}

Consider now the Hamiltonian (44) with $\alpha$-dependent hopping parameters given by
\begin{align}
\tau_n&=\4(t_1+t_2+t_3+t_4)-\4(t_1-t_2+t_3-t_4)\cos(\pi n+2\alpha)+\nn\\
&+\2(t_1-t_3)\sin(\2\pi n+\alpha)-\2(t_2-t_4)\cos(\2\pi n+\alpha),
\end{align}
with $\tau_{n+4}(\alpha)=\tau_n(\alpha)$ and $\tau_n(\alpha+2\pi)=\tau_n(\alpha)$.

From (54) we find $\tau_n(\alpha+\2\pi)=\tau_{n+1}(\alpha)$, meaning that the shift $\alpha\to\alpha+\2\pi$ corresponds
to the cyclic permutation $t_1\to t_2\to t_3\to t_4\to t_1$, hence the parameter $\alpha$ introduced by (54) performs the interpolation between the different choices of elementary cells in the chain.

The corresponding one-particle Hamiltonian is given by
\be
H=\left\lgroup\hspace*{-1.5mm}\ba{cccc}
0&\tau_1e^{+(i/4)\kappa}&0&\tau_4e^{-(i/4)\kappa}
\\
\tau_1e^{-(i/4)\kappa}&0&\tau_2e^{+(i/4)\kappa}&0
\\
0&\tau_2e^{-(i/4)\kappa}&0&\tau_3e^{+(i/4)\kappa}
\\
\tau_4e^{+(i/4)\kappa}&0&\tau_3e^{-(i/4)\kappa}&0\ea\hspace*{-1.5mm}\right\rgroup
\ee
where $\kappa=(4\ell)k$ with $4\ell$ the chain period and $-\pi\leqslant\kappa\leqslant+\pi$.

The four eigenvalues are given by
\be
2\epsilon^2_\pm(\kappa)=\tau_1^2+\tau_2^2+\tau_3^2+\tau_4^2\pm\sqrt{u+v\cos\kappa},
\ee
where
\babc
\begin{align}
u&=(\tau_1^2+\tau_2^2+\tau_3^2+\tau_4^2)^2-4\tau_1^2\tau_3^2-4\tau_2^2\tau_4^2,\\
v&=8\hh \tau_1\tau_2\tau_3\tau_4,
\end{align}
\eabc
and are depicted in Fig. 13.
\begin{center}
\includegraphics{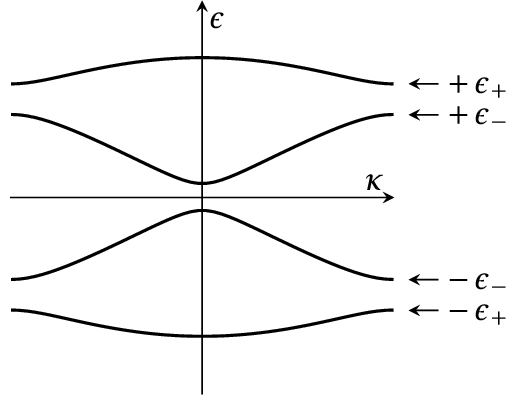}\\
{\small FIG. 13: Energy spectrum of the four-band model.}
\end{center}
\vspace*{3mm}

The spectrum is symmetric with respect to $\epsilon\to-\epsilon$ since the Hamiltonian (55) enjoys
the chiral symmetry.

Normalized eigenvectors ($s=1,2,3,4$) are given by
\be
\psi_s=\frac{1}{\sqrt{N_s}}
\left\lgroup\ba{c}
\epsilon_s\big[\tau_1\tau_2e^{+(i/2)\kappa}+\tau_3\tau_4e^{-(i/2)\kappa}\big]
\\\vspace*{-3mm}\\
\tau_2(\epsilon_s^2-\tau_4^2)e^{+(i/4)\kappa}+\tau_1\tau_3\tau_4e^{-(3i/4)\kappa}
\\\vspace*{-3mm}\\
\epsilon_s(\epsilon_s^2-\tau_1^2-\tau_4^2)
\\\vspace*{-3mm}\\
\tau_3(\epsilon_s^2-\tau_1^2)e^{-(i/4)\kappa}+\tau_1\tau_2\tau_4e^{+(3i/4)\kappa}
\ea\right\rgroup
\ee
where $N_s^{}=2\epsilon_s^2(\epsilon_s^2-\tau_1^2-\tau_4^2)(2\epsilon_s^2-\tau_1^2-\tau_2^2-\tau_3^2-\tau_4^2)>0$.

Expressions (49) -- (51) can be employed without any modification. Consequently, in order to obtain the gauge invariant
topological indices we only need the expression for $A_\kappa(\kappa,\alpha)$. Using (58) in (49a) and performing
trivial but lengthy manipulations we come to
\be
A_\kappa^\pm
=\frac{C_1}{\epsilon_\pm^2}
-\frac{C_2}{\epsilon_\pm^2-\tau_1^2-\tau_4^2}
+\frac{2C_2+C_3}{2\epsilon_\pm^2-\tau_1^2-\tau_2^2-\tau_3^2-\tau_4^2}
\ee
where
\babc
\begin{align}
4\pi C_1&=\frac{\tau_1^2\tau_3^2-\tau_2^2\tau_4^2}{\tau_1^2+\tau_2^2+\tau_3^2+\tau_4^2}\hh,\\
2\pi C_2&=\frac{\tau_1^2\tau_2^2-\tau_3^2\tau_4^2}{\tau_1^2-\tau_2^2-\tau_3^2+\tau_4^2}\hh,\\
8\pi C_3&=\frac{(\tau_1^2-\tau_2^2-\tau_3^2+\tau_4^2)(\tau_1^2+\tau_2^2-\tau_3^2-\tau_4^2)}{\tau_1^2+\tau_2^2+\tau_3^2+\tau_4^2}\hh.
\end{align}
\eabc
Mind that the connection (59) involves $\epsilon^2$ rather than $\epsilon$.

Using (56) and integrating over $\kappa$ we obtain
\begin{align}
\mu_\pm(\alpha)&\equiv\oint A_\kappa^\pm(\kappa,\alpha)d\kappa=\nn\\
&={\textstyle\frac14}{\rm sgn}(\tau_1^2\tau_3^2-\tau_2^2\tau_4^2)
-{\textstyle\frac12}{\rm sgn}(\tau_1^2\tau_2^2-\tau_3^2\tau_4^2)\mp\nn\\
&\mp\frac{(\tau_1\tau_3+\tau_2\tau_4)(\tau_1^2+\tau_2^2+\tau_3^2+\tau_4^2)}{2\pi(\tau_1\tau_3-\tau_2\tau_4)\sqrt{u+v}}\Pi\left(-\lambda\h,\frac{2v}{u+v}\right)\mp\nn\\
&\mp\frac{(\tau_1\tau_2-\tau_3\tau_4)(\tau_1^2-\tau_2^2-\tau_3^2+\tau_4^2)}{\pi(\tau_1\tau_2+\tau_3\tau_4)\sqrt{u+v}}\Pi\left(\zeta\h,\frac{2v}{u+v}\right)\pm\nn\\
&\pm\frac{\tau_1^2-\tau_2^2+\tau_3^2-\tau_4^2}{2\pi\sqrt{u+v}}K\left(\frac{2v}{u+v}\right)
\end{align}
where $K$ and $\Pi$ are the complete elliptic integrals
\babc
\begin{align}
K(x)&=\int_0^{\pi/2}\frac{d\vartheta}{\sqrt{1-x\sin^2\vartheta}}\hh,\\
\Pi(y,x)&=\int_0^{\pi/2}\frac{d\vartheta}{(1-y\sin^2\vartheta)\sqrt{1-x\sin^2\vartheta}}\hh,
\end{align}
\eabc
and
\babc
\begin{align}
\lambda&=\frac{4\tau_1\tau_2\tau_3\tau_4}{(\tau_1\tau_3-\tau_2\tau_4)^2}\hh,\\
\zeta&=\frac{4\tau_1\tau_2\tau_3\tau_4}{(\tau_1\tau_2+\tau_3\tau_4)^2}\hh.
\end{align}
\eabc

\ding{70} If the system is quarter-filled we employ $\mu_+(\alpha)$ corresponding to the lowest band $\epsilon=-\epsilon_+$
in Fig. 13. We then search for the combinations $\mu_+(\alpha_1)-\mu_+(\alpha_2)$ which (besides being gauge invariants)
would be quantized. Taking into account that the last three terms of (61) are insensitive to $\alpha\to\alpha+\pi$, we find that
the sought for quantization occurs for $|\alpha_1-\alpha_2|=\pi$. In particular, we find
\babc
\begin{align}
\sigma_2&\equiv{\rm sgn}(t_1^2t_4^2-t_2^2t_3^2)=\mu_+({\textstyle\frac12}\pi)-\mu_+({\textstyle\frac32}\pi)\hh,\\
\sigma_3&\equiv{\rm sgn}(t_1^2t_2^2-t_3^2t_4^2)=\mu_+(\pi)-\mu_+(0)\hh.
\end{align}
\eabc
These parameters are responsible for splitting off certain edge state levels from the lowest band, {\it e.g.}
$\epsilon=-(t_1^2+t_2^2)^{1/2}$ and $\epsilon=-(t_1^2+t_4^2)^{1/2}$ splitting off the lowest band in Fig. 6.
Remark that $\sigma_1\equiv{\rm sgn}(|t_1t_3|-|t_2t_4|)$ does not emerge here.
This is reasonable since $\sigma_1$ controls closing/opening of the central gap (it is closed for $t_1t_3=t_2t_4$),
which is irrelevant due to the quarter-filling.

\ding{70} For half-filling we sum up over the two lower bands. In that case we trivially come to
\begin{align}
\mu(\alpha)&\equiv\mu_+(\alpha)+\mu_-(\alpha)=\nn\\
&={\textstyle\frac12}{\rm sgn}(|\tau_1\tau_3|-|\tau_2\tau_4|)-{\rm sgn}(|\tau_1\tau_2|-|\tau_3\tau_4|)
\end{align}
where from the parameters $\sigma_{1,2,3}$ can be expressed as
\babc
\begin{align}
\sigma_1&\equiv{\rm sgn}(t_1^2t_3^2-t_2^2t_4^2)=\mu(0)-\mu({\textstyle\frac12\pi})+\mu(\pi)-\mu({\textstyle\frac32}\pi)\hh,\\
\sigma_2&\equiv{\rm sgn}(t_1^2t_4^2-t_2^2t_3^2)=\mu({\textstyle\frac12}\pi)-\mu({\textstyle\frac32}\pi)\hh,\\
\sigma_3&\equiv{\rm sgn}(t_1^2t_2^2-t_3^2t_4^2)=\mu(\pi)-\mu(0)\hh.
\end{align}
\eabc
Remark that the three combinations standing in the right hand sides of (66) are linearly independent.

\section*{5. Conclusions}

Summarizing, we have studied the issue of edge states in a tight-binding model on a finite chain with four-fold
alternated hoppings. Employing the technique developed in Ref. [7], the one-particle eigenvalue problem is solved
analytically and the wave functions are expressed in terms of the Chebyshev polynomials $U_n(\xi)$.
Energy eigenvalues of the edge states and the conditions for their formation are also found analytically.
All analytic results are confirmed by numeric calculations.

It is found that the chains with odd number of sites produce 4 phases which differ one from another with respect
to the content (presence/absence) of various edge states, while chains with even sites produce 8 different phases.

Remark also that the flat shapes of some edge state levels result from particular parameterizations.
To be clear we present the phase diagramme for $N=odd$ in Fig. 14.

Figs. 2, 3 and 4 depict the energy spectra $\epsilon_{1\div101}$ versus the parameter
$\vartheta_1$ introduced by (30). Varying $\vartheta_1$ we thus vary the ratio $t_1/t_2$ while $t_3/t_4$ is
kept constant. This corresponds to the dashed straight lines shown in Fig. 14; the lowest corresponds to $\vartheta_2<\frac14\pi$
where the phase $C$ isolates $A$ from $D$; middle one is for $\vartheta_2=\frac14\pi$, so that no intermediate
phase occurs between $A$ and $D$; and the upper one is for $\vartheta_2>\frac14\pi$ with $A$ and $D$ separated by $B$.
In these three cases the edge state levels (Figs. 2,3,4) are all flat.
\begin{center}
\includegraphics{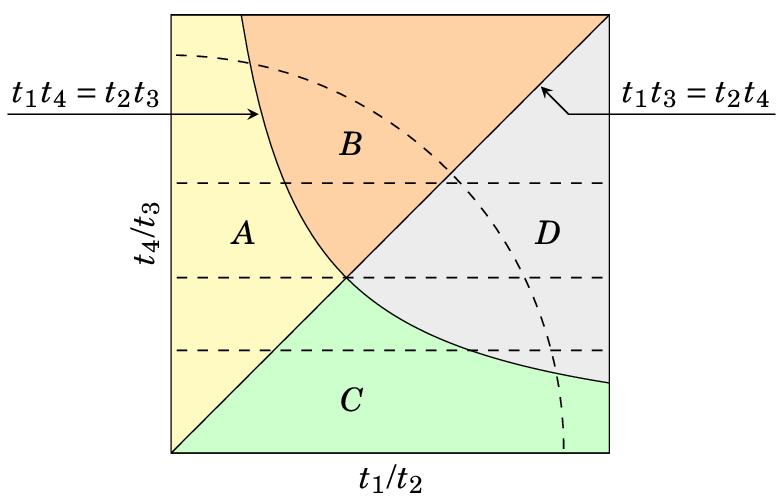}\\
{\small FIG. 14: Phase diagramme for $N=4J+1$.}
\end{center}
\vspace*{3mm}

Alternatively, one may plot the spectrum $\epsilon_{1\div101}$ versus the parameter varying along the circular path in Fig. 14.
In that case the spectrum looks as in Fig. 15 and the edge state energy levels $\epsilon=\pm(t_1^2+t_2^2)^{1/2}$
and $\epsilon=\pm(t_3^2+t_4^2)^{1/2}$ are no longer flat but interpolate between the two bands.
\begin{center}
\includegraphics{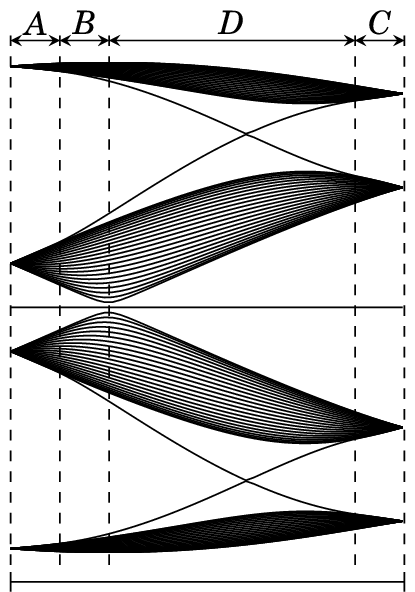}
\end{center}
\noindent
{\small FIG. 15: Energy spectrum for $N=4J+1$ versus the parameter along the circular path in Fig. 14.}
\vspace*{3mm}

Further, we have proposed the scheme for constructing gauge invariant curvature applicable to general 1D periodic chain.
In the case of the two-band model the scheme reproduces the Zak phase, while for the four-band model it recovers
the defining parameters $\sigma_{1,2,3}$ as gauge invariant topological indices.

As previously pointed out, the shift $\alpha\to\alpha+\4\pi$ in $\tau_n(\alpha)$ represents the permutation
$t_1\to t_2\to t_3\to t_4\to t_1$ which can be treated as translation of the chain in $x$-space, while the shift
of $\kappa$ represents translation in momentum space. In this light the pair $(\kappa,\alpha)$ resembles canonical
conjugate pair, while the torus might be regarded as the corresponding phase space.

\section*{Acknowledgments}

Authors are grateful to A. A. Nersesyan, G. I. Japaridze and M. Sekania for fruitful discussions.
Work was supported by Rustaveli National Science Foundation through grant FR/265/6-100/14.
D. K. appreciates support from the World Federation of Scientists.
D. K. and M. Ts. acknowledge support from the Faculty of Exact and Natural Sciences of Tbilisi
State University.

\appendix

\section{Three-Term Periodic Recursion}

We solve the recurrence relation
\be
t_n\psi_n+\epsilon\psi_{n+1}+t_{n+1}\psi_{n+2}=0
\ee
provided the coefficients are periodic $t_{n+\omega}=t_n$.

For this purpose we employ the results of Ref. [7].
Let $p_n(x)$ be a sequence of polynomials set by the recursion
\be
p_n(x)=(x+b_{n-1})p_{n-1}(x)-a_{n-1}p_{n-2}(x)
\ee
supplied by the boundary conditions $p_{-1}=0$ and $p_0=1$,
and the periodic coefficients $a_{n+\omega}=a_n$ and $b_{n+\omega}=b_n$.

As shown in Ref. [7] the polynomial $p_{\omega-1}$ divides $p_{2\omega-1}$
\be
q(x)=\frac{p_{2\omega-1}(x)}{p_{\omega-1}(x)}
\ee
where $q(x)$ is of order of $\omega$.

Then the solution to (A2) appears as
\be
p_{\omega n+s}=a^{n-1}p_{s+\omega}U_{n-1}(q/2a)-a^np_sU_{n-2}(q/2a)
\ee
where $a^2\equiv a_1a_2\cdots a_\omega$.

We proceed to bring (A1) to the form (A2).
Introducing $p_n\equiv(t_1\cdots t_n)\psi_{n+1}$ we rewrite (A1) as
\be
p_{n+1}=-\epsilon p_n-t^2{\hspace*{-1.6mm}}_n p_{n-1}
\ee
and the boundary condition $\psi_0=0$ takes the form $p_{-1}=0$.

We thus come to the recursion set by (A2) with $a_n=t^2{\hspace*{-1.6mm}}_n$
and $b_n=0$.

\ding{70} For the two-band model the periodicity is $\omega=2$, and $q(\epsilon)$ set by (A3) appears as
\be
q(\epsilon)=\epsilon^2-t^2_1-t^2_2.
\ee

Using (A4) we solve $p_n$ and rewriting in terms of $\psi_n$ find
\be
\psi_{2n+s}(\epsilon)=\psi_{s+2}(\epsilon)U_{n-1}(\xi)-\psi_s(\epsilon)U_{n-2}(\xi),
\ee
where
\be
\xi\equiv\frac{\epsilon^2-t^2_1-t^2_2}{2t_1t_2}.
\ee

Writing out (A7) for $s=1,2$ we find
\babc
\begin{align}
\psi_{2n+1}&=\psi_3U_{n-1}(\xi)-\psi_1U_{n-2}(\xi),\\
\psi_{2n+2}&=\psi_4U_{n-1}(\xi)-\psi_2U_{n-2}(\xi).
\end{align}
\eabc

Employing the recursion (A1) we express $\psi_{2,3,4}$ in terms of $\psi_1$ and rewrite (A9) as (4).

\ding{70} For the four-band model the periodicity is $\omega=4$, and $q(\epsilon)$ set by (A3) appears as
\be
q(\epsilon)=\epsilon^4-(t_1^2+t_2^2+t_3^2+t_4^2)\epsilon^2+t_1^2t_3^2+t_2^2t_4^2.
\ee

Using (A4) we solve $p_n$ and rewriting in terms of $\psi_n$ find
\be
\psi_{4n+s}(\epsilon)=\psi_{s+4}(\epsilon)U_{n-1}(\xi)-\psi_s(\epsilon)U_{n-2}(\xi),
\ee
where
\be
\xi\equiv\frac{\epsilon^4-(t_1^2+t_2^2+t_3^2+t_4^2)\epsilon^2+t_1^2t_3^2+t_2^2t_4^2}{2t_1t_2t_3t_4}.
\ee

Writing out (A11) for $s=1,2,3,4$ we find
\babc
\begin{align}
\psi_{4n+1}&=\psi_5U_{n-1}(\xi)-\psi_1U_{n-2}(\xi),\\
\psi_{4n+2}&=\psi_6U_{n-1}(\xi)-\psi_2U_{n-2}(\xi),\\
\psi_{4n+3}&=\psi_7U_{n-1}(\xi)-\psi_3U_{n-2}(\xi),\\
\psi_{4n+4}&=\psi_8U_{n-1}(\xi)-\psi_4U_{n-2}(\xi).
\end{align}
\eabc

Employing (A1) we express $\psi_{2\div8}$ via $\psi_1$ and rewrite (A13)
as (22). In so doing we use $U_{n+1}(x)=2xU_n(x)-U_{n-1}(x)$.

\section{$N=4J+1$}

We search for the edge states provided $J\to\infty$. For this purpose we study the equation (26a)
\be
\epsilon\Bigg[\frac{U_J(\xi)}{U_{J-1}(\xi)}+\frac{t_1t_4}{t_2t_3}\Bigg]=0.
\ee

Since the edge states emerge only when $|\xi|>1$, we analyze
the equation (B1) for $\xi>1$ and $\xi<-1$.

\ding{70} $\xi>1$. In this case we have $U_J(\xi)>0$, hence the only solution to (B1) is given by $\epsilon=0$.
Using this in (23) we find
\be
\xi=\frac12\Bigg(\frac{t_1t_3}{t_2t_4}+\frac{t_2t_4}{t_1t_3}\Bigg)\geqslant1
\ee
{\it i.e.} we have $\xi>1$ except when $t_1t_3=t_2t_4$ where $\xi=1$.

We are now about to use (B2) in (22) and (25) involving the quantities $U_n(\xi)$.
Here we use the relation
\be
U_n\Bigg[\frac12\Bigg(x+\frac{1}{x}\Bigg)\Bigg]=\frac{1-x^{2n+2}}{x^n(1-x^2)}
\ee
which in fact is the same (10) and allows to calculate $U_n(\xi)$ in the exact way.
For $t_1t_3<t_2t_4$ we use (22) and come to (27), while for $t_1t_3>t_2t_4$ we use (25) and come to (28).

\ding{70} $\xi<-1$. In this case we put $\xi=-\hh\cosh z$ and using (10) write the ratio of two polynomials
standing in (B1) as
\be
\frac{U_J(\xi)}{U_{J-1}(\xi)}=-\frac{\sinh[(J+1)z]}{\sinh[Jz]}.
\ee

Taking $J\to\infty$ we put $z>0$ (the same final result occurs for $z<0$) and come to
\be
\frac{U_J(\xi)}{U_{J-1}(\xi)}\to-\hh e^z.
\ee
Using this in (B1) we obtain
\be
e^z=\frac{t_1t_4}{t_2t_3}.
\ee
Provided $z$ is taken to be positive, the last relation implies that the case under consideration can be realized only if $t_1t_4>t_2t_3$.

Substituting (B6) into $\xi=-\cosh z$ we rewrite the equation (23) as
$(\epsilon^2-t_1^2-t_2^2)(\epsilon^2-t_3^2-t_4^2)=0$ producing four levels
\babc
\begin{align}
\epsilon&=\pm(t_1^2+t_2^2)^{1/2},\\
\epsilon&=\pm(t_3^2+t_4^2)^{1/2}.
\end{align}
\eabc
For (B7a) we use (22) and come to (29a), while for (B7b) we use (25) and come to (29b).

\section{$N=4J+2$}

We study the secular equation (26b) in the limit of $J\to\infty$ and consider the cases $|\xi|>1$.

\ding{70} $\xi>1$. Taking $\xi=\cosh z$ we write the equation (26b) as
\be
\frac{\epsilon^2-t_1^2}{t_1t_2}\hh\frac{\sinh[(J+1)z]}{\sinh[Jz]}+\frac{t_4}{t_3}=0.
\ee
Assuming $z>0$, and taking the limit $J\to\infty$ this leads to
\be
e^z=-\frac{t_4}{t_3}\frac{t_1t_2}{\epsilon^2-t_1^2}
\ee
Substituting (C2) into $\xi=\cosh z$ and combining with (23) we come up to the following three equations
\babc
\begin{align}
\epsilon^2&=0,\\
\epsilon^2&=t_1^2+t_2^2,\\
\epsilon^2&=t_1^2+t_4^2.
\end{align}
\eabc

Last two options are controversial since the right hand side of (C2) becomes negative.
Using (C3a) in (C2) we find
\be
e^z=\frac{t_2t_4}{t_1t_3},
\ee
{\it i.e.} provided $z>0$, we may have $\xi>1$ only for $t_1t_3<t_2t_4$ with $\epsilon^2=0$.
The fact that the eigenvalue occurs as $\epsilon^2=0$ signifies the energy level with $\epsilon=0$ is doubly
degenerated. One of the two wave functions is obtained by assuming $\psi_1$ is finite. In that case we use
(22) and come to (31a). The other is obtained by assuming $\psi_N$ is finite. In that case we use (25)
which leads to (31b).

\ding{70} $\xi<-1$. We put $\xi=-\cosh z$ and rewrite (C1) as
\be
\frac{\epsilon^2-t_1^2}{t_1t_2}\hh\frac{\sinh[(J+1)z]}{\sinh[Jz]}-\frac{t_4}{t_3}=0
\ee
Assuming $z>0$, and taking the limit $J\to\infty$ we obtain
\be
e^z=\frac{t_4}{t_3}\frac{t_1t_2}{\epsilon^2-t_1^2}.
\ee

Substituting this into $\xi=-\cosh z$ and combining with (23) we come to the same three options given by (C3).
In this case the first option is controversial and we study the last two
\babc
\begin{align}
\epsilon^2&=t_1^2+t_2^2\5\Longrightarrow\5e^z=t_1t_4/t_2t_3,\\
\epsilon^2&=t_1^2+t_4^2\5\Longrightarrow\5e^z=t_1t_2/t_3t_4.
\end{align}
\eabc
which may occur for $t_1t_4>t_2t_3$ and $t_1t_2>t_3t_4$, respectively.

For (C7a) we use (22) and come to (32). For (C7b) we use (25) which gives (33).

\section{$N=4J+3$}

Secular equation (26c) breaks into the following three
\babc
\begin{align}
&\epsilon=0,\\
&\epsilon^2=t_1^2+t_2^2,\\
&U_J(\xi)=0.
\end{align}
\eabc
The roots of Chebyshev polynomials are all located in the segment $(-1,+1)$, hence (D1c) leads to $|\xi|<1$, {\it i.e.} to bulk states.

\ding{70} Using $\epsilon=0$ in (23) we find $\xi=\frac12(t_1t_3/t_2t_4)+\frac12(t_2t_4/t_1t_3)$
and we consider the two options $t_1t_3<t_2t_4$ and $t_1t_3>t_2t_4$. In the first case we use (22)
and come to (34), while for the other we use (25) and come to (35).

\ding{70} Using $\epsilon^2=t_1^2+t_2^2$ in (23) we obtain
\be
\xi=-\hh\frac{t_1^2t_4^2+t_2^2t_3^2}{2t_1t_2t_3t_4}\leqslant-1
\ee
and we have two options $t_1t_4>t_2t_3$ and $t_1t_4>t_2t_3$. In the first case we use (22) and come to (36).
For the other we use (25) and come to (37).

\section{$N=4J+4$}

\ding{70} $\xi>1$. Taking $\xi=\cosh z$ the equation (26d) reads
\be
\frac{\sinh[(J+2)z]}{\sinh[(J+1)z]}+\frac{t_4}{t_3}\hh\frac{\epsilon^2-t_2^2}{t_1t_2}=0.
\ee
Assuming $z>0$ we take the limit $J\to\infty$ and obtain
\be
e^z=-\hh\frac{t_4}{t_3}\hh\frac{\epsilon^2-t_2^2}{t_1t_2}.
\ee

Using this in $\xi=\cosh z$ and combining with (23) we find
\babc
\begin{align}
\epsilon^2&=0,\\
\epsilon^2&=t_1^2+t_2^2,\\
\epsilon^2&=t_2^2+t_3^2.
\end{align}
\eabc

(E3b) and (E3c) controversial since the right hand side of (E2) becomes negative.
Using (E3a) in (E2) we find
\be
e^z=\frac{t_2t_4}{t_1t_3},
\ee
{\it i.e.} provided $z>0$, we may have $\xi>1$ only if $t_1t_3<t_2t_4$ with $\epsilon^2=0$.
The fact that the eigenvalue occurs as $\epsilon^2=0$ implies the energy level with $\epsilon=0$ is doubly
degenerated. One of the two is obtained using (22) and leads to (38a). The other one is obtained using (25) and appears as (38b).

\ding{70} $\xi<-1$. Taking $\xi=-\cosh z$, the secular equation (26d) reads
\be
\frac{\sinh[(J+2)z]}{\sinh[(J+1)z]}-\frac{t_4}{t_3}\hh\frac{\epsilon^2-t_2^2}{t_1t_2}=0.
\ee
Assuming $z>0$, and taking the limit $J\to\infty$ we obtain
\be
e^z=\frac{t_4}{t_3}\hh\frac{\epsilon^2-t_2^2}{t_1t_2}.
\ee

Substituting this into $\xi=-\cosh z$ and combining with (23), we obtain the same three options (E3).
First option is controversial, and we study the last two ones
\babc
\begin{align}
\epsilon^2&=t_1^2+t_2^2\5\Longrightarrow\5e^z=\frac{t_1t_4}{t_2t_3},\\
\epsilon^2&=t_2^2+t_3^2\5\Longrightarrow\5e^z=\frac{t_3t_4}{t_1t_2},
\end{align}
\eabc
hence the first one occurs if $t_1t_4>t_2t_3$, and the second one occurs for $t_1t_4<t_2t_3$.
In the first case we use (22) find (39), while in the second one we use (25) and come to (40).

\end{document}